# A Study of VLSI Technology, Wafers and Impact on Nanotechnology

Mrs. Kiran Gupta* Member IEEE,
*Asst.Professor, DSCE ,B'lore (INDIA). E-mail:18.kiran@gmail.com
Dr. T R.Gopalakrishnan.Nair, Senior Member IEEE,
Director Research and Industry Incubation center, DSCE, Blore. E-mail: trgnairi@yahoomail.com

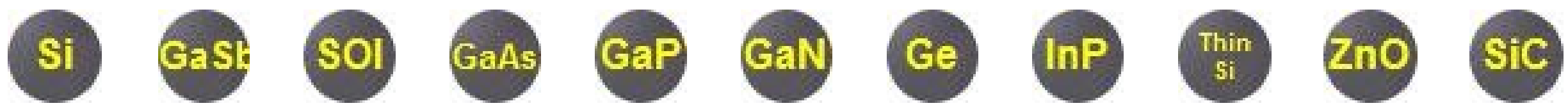

***Abstract -*** **This paper presents a detailed study of the present VLSI technological aspects, importance and their replacement or combination with the Nanotechnology in the VLSI world of silicon semiconductors. Here authors bring out the nanotechnology in Silicon world which invariably means shrinking geometry of CMOS devices to nano scale. This also refers to a new world of nanotechnology where chemists are working in manufacturing of carbon nanotubes , nano devices of varius materials of nano dimensions without even knowing how this could change the whole world of Si and CMOS technology and the world we live in.**

## I. Introduction

The VLSI technology means 10s of millions of CMOS transistors in microns on a silicon wafer of a few cm dimensions. GaAs devices also are found in a few applications but Si dominates the VLSI chips. The Moore's law continues to hold good with the continuous advances in the VLSI technology and design issues. Also different materials are replacing the conventional silicon wafers and aluminum metal interconnects to achieve more density per chip to cope up with the miniaturization in the technology and it is believed that their might be a dead end to the CMOS technology in future if we try to keep on going with this trend [1]. Silicon wafers have been replaced with the Silicon-on-Insulator and low k-dielectric cost of chip masks and next-generation fabrication plants totally replacing today's foundries and copper interconnects leading to the production of CMOS process with effective line widths of less than 120 nm. In spite of advances this industry and research, a threat to economic issues holds as the continuing scaling of Silicon transistors brings about a new, exciting era of nano-scale technology to silicon foundries.

## II. Nanotechnology and CAEN

Chemically assembled electronic Nanotechnology (CAEN), a form of electronic nanotechnology (EN), that uses Self-alignment to construct electronic circuits out of Nanometer-scale devices. CAEN devices are a promising alternative to CMOS-based devices; particularly CMOS based reconfigurable devices [2]. Following International Technology Roadmaps for Semiconductor (ITRS), advanced silicon foundries are manufacturing a full spectrum of integrated-circuit products down to 90-nm technology node today. A few companies like IBM and INTEL have launched a few chips using 65 nm technologies, eg. Quad core processor using 65nm [7] . At this stage, many of the device characteristics are no longer a straightforward extension of past generations. Scaling is beyond simple shrinking of 3D physical dimensions of devices; it involves the changing, or 'straining', of the atomic spacing to alter the Silicon electronic properties for better performance.

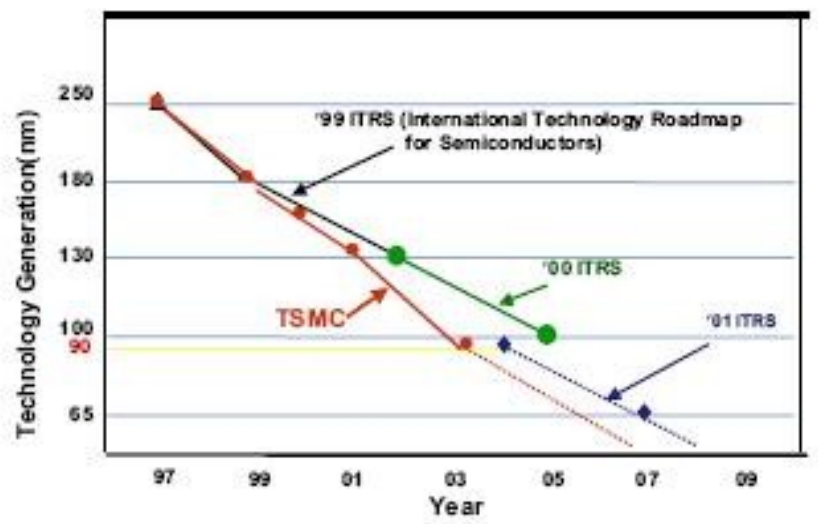

Fig. 1 ITRS roadmap and offering date of Foundry technology

The research somewhat halts here to look forward to a new world of Nanotechnology to replace Silicon and metallic interconnects with carbon nano devices and carbon nanotubes. CAEN can be combined with reconfigurable computing to create useful computational devices which are both economical to construct and which can tolerate a high-degree of variability. The result will be a high-density low-power technology with inherently lower fabrication costs than its CMOS counterpart.

## III. WAFERS in Nano CHIPS

A study of the documents from the research labs and marketing organization around the world reveal that concept of Silicon wafers is undergoing a change slowly with the SOI and other chips. If CMOS Nanotechnology is to be replaced by CAEN, the availability of wafer spectrum for research wi; be as wide as follows:

*A. SOI Wafers*

**SOI: silicon on insulator**
2" SOIP(100) 500anstrom Si layer
4" SOINor P (100) Device 2-4 m

SOI layer diameter: 200mm
crystal orientation: <100> 4 deg off-axis
Dopant: N type (Phosphor)
SOI layer thickness: 3.0um
SOI layer resistivity: 5-20 ohm/cm
Buried Oxide: 1.0-1.5um
Base Wafer Dopant: N type (phosphor)
resistivity: 10-20 ohm/cm
thickness: 725 um

*B. Si and Ga special wafers*

Silicon Wafer special
DSP Silicon wafers P-type 1-0-0 Boron> 700 m-3000mm
(i) Thermal oxide layer (50.8mm-300mm) –1K-10 K of oxide
(ii) NITRIDE wafers: 4" N/Ph (100)

## III. Nano wires and Interconnects

Stochastically assembled nanoscale architectures have the potential to achieve device densities 100 times greater than today's CMOS. A key challenge facing nanotechnologies is controlling parallel sets of nanowires (NWs), such as those in crossbars, using a moderate number mesoscale wires. There are three methods to do this. The first is based on NW differentiation during manufacture, the second makes random connections between NWs and mesoscale wires, and the third, a mask-based approach, interposes high-K between dielectric region between NWs and mesoscale wires. Each of these addressing schemes involves a stochastic step in their implementation. The third approach is mask-based approach and shows that, when compared to the other two schemes, a large number of mesoscale control wires are necessary for its realization.

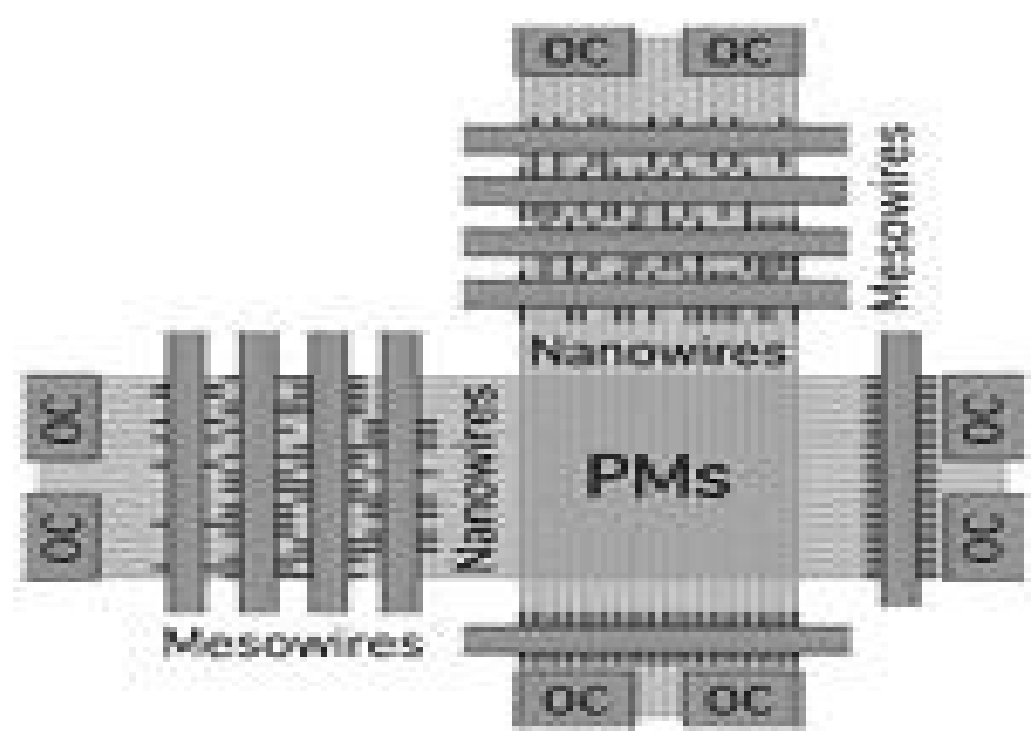

Fig. 2. Nano Interconnect

### IV. NANOTRANSISORS

It is required to unerstand and study on the carrier transport theory to understand the electron conduction behavior in transistors smaller than 20-nm. In nano-scale transistors, the number of atoms in the active region is finite; the nature of random distribution of atoms in the active region causes fluctuation in device property and deteriorates the design margins for integration. Nonetheless, scientists and engineers in Foundry master the variability in device

GaAs Wafers:
2" GAAs N(100) 350 M DSP
4" GaAs Si (100) 600M DSP

The materials that belong to the III and V group of periodic table and find their prominent place in the VLSI chips are GaP, GaSB, InAs, InSB and InP. The other wafers used are Quartz wafer, Pyrex wafer, apires, Silicon Nitride wafers, Germanium wafers, Silicon Ga AS, Fused silica, Float zone silica, undoped silica characteristics, the integration level of nano-scale transistors continues to rise yearly, and has exceeded 100-million transistors on a chip. Foundries have invested heavily in R&D and new facility to perfect the art of producing such nano-scale devices with extremely low defect density. The precise control of each step in the manufacturing process flow ensures repeatability and uniformity. While devices are being miniaturized, the wafer size is increased to improve productivity. The success of producing such nano-scale devices in an ever-increasing integration level on a bigger size wafer and at lower defect density is the 'quiet' side of the nano-electronics evolution. A seemly antiquated, 40-years old, silicon technology is now being threatened.

## VI. CONCLUSION

The goal was to study the various materials used in VLSI technology. The study reveals that present scaling of the CMOS technology to nano dimensions will have to limit at some point and make further scaling may be impossible while retaining all the electrical characteristics of the devices. This may be replaced by chemically engineered nano devices with all together different concepts leading to dramatic size reduction. This further may lead to reinvestment in nano-foundries to meet the challenges laid.

## ACKNOWLEDGMENTS

The authors would like to thank the authorities of DSCE Bangalore (India) for the support given and thanks to Dr. A. Sreenivasan , Director , PG studies in Engineering, DSCE for his valuable suggestions and support.